\documentclass{article}
\usepackage{amsmath, amssymb, graphicx, setspace}
\usepackage[margin=2.0cm]{geometry}

\newcommand{\mathsym}[1]{{}}
\newcommand{\unicode}[1]{{}}

\begin{document}

\title{Position eigenstates via application of an operator on the vacuum}
\author{H{\' e}ctor Manuel Moya-Cessa and Francisco Soto Eguibar\\
\small{Instituto Nacional de Astrof\'{i}sica, \'{O}ptica y Electr\'{o}nica INAOE} \\
\small{Luis Enrique Erro 1, Santa Mar\'{i}a Tonantzintla, San Andr\'{e}s Cholula, Puebla, 72840 M\'{e}xico}}
\date{}
\maketitle

\begin{abstract}The squeezed states are states of minimum uncertainty, but unlike the coherent states, in which the uncertainty in the position and
the momentum are the same, these allow to reduce the uncertainty, either in the position or in the momentum, while maintaining the principle of uncertainty
in its minimum. It seems that this property of the squeezed states would allow you to get the position eigenstates as a limit case of them, doing
null the uncertainty in the position and infinite at the momentum. However, there are two equivalent ways to define the squeezed states, which lead
to different expressions for the limit states. In this work, we analyze these two definitions of the squeezed states and show the advantages and
disadvantages of using such definition to find the position eigenstates. With this idea in mind, but leaving aside the definitions of the squeezed
states, we find an operator applied to the vacuum that gives us the position eigenstates. We also analyze some properties of the squeezed states,
based on the new expressions obtained for the eigenstates of the position.
\end{abstract}

\section{Introduction}
The easiest to understand and to manipulate, and the most natural states of the quantum harmonic oscillator are the number states \(|n\rangle\). Number
states are eigenstates of the Hamiltonian and, of course, are also eigenstates of the number operator \(\hat{n}=\hat{a}^{\dagger }\hat{a}\), where
\(\hat{a}^{\dagger }\) and \(\hat{a}\) are the well known creation and annihilation operators, respectively. However, for any \(n\), no matter how
big, the mean field is zero; i.e., \(\langle n|\hat{E}_x|n\rangle =0\), and we know that a classical field changes sinusoidally in time in each point
of space; thus, these states can not be associated with classical fields [1, 2].

In the first years of the sixties of the past century, Glauber [3] and Sudarshan [4] introduced the coherent states, and it has been shown that these
states are the most classical ones. Coherent states are denoted as \(|\alpha \rangle\), and one way to define them is as eigenstates of the annihilation
operator; that is, \(\hat{a}|\alpha \rangle =\alpha |\alpha \rangle\). An equivalent definition is obtained applying the Glauber displacement operator
\(\hat{D}(\alpha )=\exp \left(\alpha \hat{a}^{\dagger }-\alpha ^*\hat{a}\right)\) to the vacuum: \(|\alpha \rangle =\hat{D}(\alpha )|0\rangle\);
we see then coherent states as vacuum displaced states. Coherent states also have the very important property that they minimize the uncertainty
relation for the two orthogonal field quadratures with equal uncertainties in each quadrature [1, 2].

Since then, other states have been introduced. In particular, squeezed states [5] have attracted a great deal of attention over the years because
their properties allow to reduce the uncertainties either of the position or momentum, while still keeping the uncertainty principle to its minimum.
Because of this, they belong to a special class of states named minimum uncertainty states. Once produced, for instance as electromagnetic fields
in cavities, they may be monitored via two level atoms in order to check, or measure, that such states have been indeed generated [6, 7].

Based on the above properties, we can think on the eigenstates of the position as limit cases of the squeezed states. As the squeezed states are
minimum uncertainty states, we can reduced to zero the uncertainty in the position, while the uncertainty in the momentum goes to infinity, so that
we keep the uncertainty principle to its minimum. Of course, there is also the option to reduce to zero the uncertainty in the momentum, while the
position gets completely undefined, obtaining that way the possibility to define momentum eigenstates. In Sections 1 and 2, we analyze the possibility
of define the position eigenstates as the limit of extreme squeezing of the squeezed states. In what follows, we will use a unit system such that
\(\hbar =m=\omega =1\).

There are two equivalent forms to define the squeezed states. In the first one, introduced by Yuen [8], squeezed states are obtained from the vacuum
as
\begin{equation}
|\alpha ;r\rangle =\hat{S}(r)\hat{D}(\alpha )|0\rangle =\hat{S}(r)|\alpha \rangle ,
\end{equation}
where
\begin{equation}
\hat{S}(r)=\exp \left[\left(\hat{a}^2-\hat{a}^{\dagger^{2}}\right)r/2\right]
\end{equation}
is the so-called squeeze operator. In this view, squeezed states are created displacing the vacuum, and after, squeezing it. Note that when the squeeze
parameter \(r\) is zero, the squeezed states reduce to the coherent states. In this work, we will consider only real squeeze parameters, as that
is enough for our intentions.

In the way followed by Caves [9], the vacuum is squeezed and the resulting state is then displaced; that means, that in this approach
\begin{equation}
|\alpha ';r'\rangle =\hat{D}\left(\alpha '\right)\hat{S}\left(r'\right)|0\rangle .
\end{equation}
Both definitons of the squeezed states agree when the squeeze factor is the same, \(r'=r\), and when the modified amplitude \(\alpha '\) of the Caves
approach is given by
\begin{equation}
\alpha '=\mu \alpha -\nu \alpha ^*,
\end{equation}
being
\begin{equation}
\mu =\cosh \, r
\end{equation}
and
\begin{equation}
\nu =\sinh \, r.
\end{equation}

To analyze the uncertainties in the position and in the momentum of the squeezed states, we introduce, following Loudon and Knight [5], the quadrature
operators
\begin{equation}
\hat{X}=\frac{\hat{a}+\hat{a}^{\dagger }}{2}=\frac{\hat{x}}{\sqrt{2}}
\end{equation}
and
\begin{equation}
\hat{Y}=\frac{\hat{a}-\hat{a}^{\dagger }}{2i}=\frac{\hat{p}}{\sqrt{2}}
\end{equation}
where \(\hat{x}\) is the position operator and \(\hat{p}\) the momentum operator. Note that the quadrature operators are essentially the position and momentum operators; this definition just provides us with two operators that have the same dimensions.

In order to show that really the squeezed states are minimum uncertainty states, we need to calculate the expected values in the squeezed state (1) of the quadrature operators (7) and (8), and its squares. Using (7) and (1), we get
\begin{equation}
\langle \alpha ;r|\hat{X}|\alpha ;r\rangle =\frac{1}{2}\langle \alpha |\hat{S}^{\dagger }(r)\frac{\hat{a}+\hat{a}^{\dagger }}{2}\hat{S}(r)|\alpha\rangle .
\end{equation}
The action of the squeeze operator on the creation and annihilation operators is obtained using the Hadamard's lemma [10, 11],
\begin{equation}
\hat{S}^{\dagger }(r)\hat{a}\hat{S}(r)=\mu \hat{a}-\nu \hat{a}^{\dagger },\text{          }\hat{S}^{\dagger }(r)\hat{a}^{\dagger }\hat{S}(r)=\mu \hat{a}^{\dagger }-\nu \hat{a},
\end{equation}
such that
\begin{equation}
\hat{S}^{\dagger }(r)\frac{\hat{a}+\hat{a}^{\dagger }}{2}\hat{S}(r)=e^{-r}\hat{X}.
\end{equation}
Therefore, as $\hat{a}|\alpha \rangle =\alpha |\alpha \rangle$ and \(\langle \alpha |\hat{a}^{\dagger }=\langle \alpha |\alpha ^*\), it is easy to see that
\begin{equation}
\langle \alpha ;r|\hat{X}|\alpha ;r\rangle =e^{-r}\frac{\alpha +\alpha ^*}{2},
\end{equation}
and that
\begin{equation}
\langle \alpha ;r|\hat{X}^2|\alpha ;r\rangle =e^{-2r}\frac{1+2\left| \alpha \right| ^2+\alpha ^2+\alpha ^{*^{2}}}{4}.
\end{equation}
So, we obtain for the uncertainty in the quadrature operator \(\hat{X}\),
\begin{equation}
\Delta \, X\equiv \sqrt{\langle \alpha ;r|\hat{X}^2|\alpha ;r\rangle -\langle \alpha ;r|\hat{X}|\alpha ;r\rangle ^2}=\frac{e^{-r}}{2}.
\end{equation}
Proceeding in exactly the same way for the quadrature operator \(\hat{Y}\), we obtain
\begin{equation}
\Delta \, Y\equiv \sqrt{\langle \alpha ;r|\hat{Y}^2|\alpha ;r\rangle -\langle \alpha ;r|\hat{Y}|\alpha ;r\rangle ^2}=\frac{e^r}{2}.
\end{equation}
As we already said, we can then think in the position eigenstates and in the momentum eigenstates as limit cases of squeezed states. Indeed, when the squeeze parameter \(r\) goes to infinity, the uncertainty in the position goes to zero, and the momentum is completely undetermined. Of course, when the squeeze parameter goes to minus infinity, we have the inverse situation, and we can think in define that way the momentum eigenstates. In the two following sections, we use the Yen and the Caves definitions of the squeezed states to test this hypothesis.

\section{A first attempt {\` a} la Yuen}
From equation (14) above, we can see that in the limit $r\to \infty$ the uncertainty for position vanishes and so a position eigenstate should
be obtained (from now on, we consider \(\alpha\) real),
\begin{equation}
\lim_{r\to \infty } |\frac{x}{\sqrt{2}};r\rangle \to |x\rangle _p.
\end{equation}
We have written a sub index \(p\) in the position eigenstate in order to emphasis that fact.
Following the Yuen definition \(|\alpha ;r\rangle =\hat{S}(r)\hat{D}(\alpha )|0\rangle =\hat{S}(r)|\alpha \rangle\), so
\begin{equation}
|\frac{x}{\sqrt{2}};r\rangle =\hat{S}(r)\hat{D}\left(\frac{x}{\sqrt{2}}\right)|0\rangle =\hat{S}(r)|\frac{x}{\sqrt{2}}\rangle .
\end{equation}
We now write the squeeze operator as [12]
\begin{equation}
\hat{S}(r)=\frac{1}{\sqrt{\mu }}
e^{-\frac{\nu }{2\mu }\hat{a}^{\dagger^{2} }}
\frac{1}{\mu ^{\hat{a}^{\dagger }\hat{a}}}e^{\frac{\nu }{2\mu }\hat{a}^2},
\end{equation}
where, as we already said, \(\mu =\cosh \, r\) and \(\nu =\sinh \, r\). So,
\begin{equation}
|\frac{x}{\sqrt{2}};r\rangle =\frac{1}{\sqrt{\mu }}e^{-\frac{\nu }{2\mu }\hat{a}^{\dagger^{2} }}\frac{1}{\mu ^{\hat{a}^{\dagger }\hat{a}}}e^{\frac{\nu}{2\mu }\hat{a}^2}|\frac{x}{\sqrt{2}}\rangle .
\end{equation}
Now, we develop the first operator (from right to left) in power series, we use the definition of the coherent states, \(\hat{a}|\alpha \rangle =\alpha
|\alpha \rangle\), and the action of the number operator over the number states \(\left(\hat{a}^{\dagger }\hat{a}|n\rangle =\hat{n}|n\rangle =n|n\rangle
\right)\), to obtain
\begin{equation}
|\frac{x}{\sqrt{2}};r\rangle =\frac{1}{\sqrt{\mu }}e^{-\frac{\nu }{2\mu }\hat{a}^{\dagger^{2} }}\left(\frac{1}{\mu }\right)^{\hat{a}^{\dagger }\hat{a}}\sum
_{n=0}^{\infty } \left(\frac{x}{\sqrt{2}}\right)^n\frac{1}{\sqrt{n!}}|n\rangle =\frac{1}{\sqrt{\mu }}e^{-\frac{\nu }{2\mu }\hat{a}^{\dagger^{2} }}\sum
_{n=0}^{\infty } \left(\frac{x}{\sqrt{2}}\right)^n\frac{1}{\sqrt{n!}}\left(\frac{1}{\mu }\right)^n|n\rangle .
\end{equation}
As \(r\to \infty\), \(\frac{1}{\mu }=\frac{1}{\cosh \, r}\to 0\), which means that the only term that survives from the sum is \(n=0\), and then
\begin{equation}
|x\rangle _p\propto e^{-\frac{\nu }{2\mu }\hat{a}^{\dagger^{2} }}|0\rangle
\end{equation}
that would give an approximation for how to obtain a position eigenstate from the vacuum. However, note that the above expression does not depend on \(x\) and therefore can not be correct.

\section{A second attempt {\` a} la Caves}
We now squeeze the vacuum and after we displace it. Thus, in this case,
\begin{equation}
|x\rangle _p=\lim_{r\to \infty } |\frac{x}{\sqrt{2}};r\rangle =\lim_{r\to \infty } \hat{D}\left(\frac{x}{\sqrt{2}}\right)\hat{S}(r)|0\rangle .
\end{equation}
We use again expression \(\hat{S}(r)=\exp \left(-\frac{\nu }{2\mu }\hat{a}^{\dagger^{2} }\right)\left(\frac{1}{\mu }\right)^{\hat{n}+\frac{1}{2}}\exp \left(\frac{\nu}{2\mu }\hat{a}^2\right)\text{}\) for the squeeze operator [12], where $\mu $ and $\nu $ are defined in (5) and (6), and we write the displacement operator
as \(\hat{D}(\alpha )=\exp \left(\frac{\left|\alpha |^2\right.}{2}\right)\exp \left(-\alpha ^* \hat{a}\right)\exp \left(\alpha \hat{a}^{\dagger }\right)\) [12], to obtain
\begin{equation}
|\frac{x}{\sqrt{2}};r\rangle =\exp \left(\frac{x^2}{4}\right)\exp \left(-\frac{x}{\sqrt{2}}\, \hat{a}\right)\exp \left(\frac{x}{\sqrt{2}}\hat{a}^{\dagger
}\right)\exp \left(-\frac{\nu }{2\mu }\hat{a}^{\dagger ^{2}}\right)\left(\frac{1}{\mu }\right)^{\hat{n}+\frac{1}{2}}\exp \left(\frac{\nu }{2\mu }a^2\right)|0\rangle.
\end{equation}
As \(\hat{a}|0\rangle =0\) and \(\hat{a}^{\dagger }\hat{a}|0\rangle =\hat{n}|0\rangle =0\), we cast the previous formula as
\begin{equation}
|\frac{x}{\sqrt{2}};r\rangle =\frac{1}{\sqrt{\mu }}\exp \left(\frac{x^2}{4}\right)\exp \left(-\frac{x}{\sqrt{2}}\, \hat{a}\right)\exp \left(\frac{x}{\sqrt{2}}\hat{a}^{\dagger
}\right)\exp \left(-\frac{\nu }{2\mu }\hat{a}^{\dagger ^{2}}\right)|0\rangle .
\end{equation}
Inserting two times the identity operator, written as \(\hat{I}=\exp \left(\frac{x}{\sqrt{2}}\, \hat{a}\right)\exp \left(-\frac{x}{\sqrt{2}}\, \hat{a}\right)\),
we get
\begin{equation}
|\frac{x}{\sqrt{2}};r\rangle =\frac{1}{\sqrt{\mu }}
e^ {\frac{x^2}{4}}
e^ {-\frac{x}{\sqrt{2}}\, \hat{a}}
e^ {\frac{x}{\sqrt{2}}\hat{a}^{\dagger}}
e^ {\frac{x}{\sqrt{2}}\, \hat{a}}
e^ {-\frac{x}{\sqrt{2}}\, \hat{a}}
e^ {-\frac{\nu }{2\mu }\hat{a}^{\dagger^{2} }}
e^ {\frac{x}{\sqrt{2}}\, \hat{a}}
e^ {-\frac{x}{\sqrt{2}}\, \hat{a}}
|0\rangle .
\end{equation}
It is clear that \(\exp \left(-\frac{x}{\sqrt{2}}\, \hat{a}\right)|0\rangle =|0\rangle\), and using the Hadamard\'{ }s lemma [10], it is easy to prove that
\begin{equation}
\exp \left(-\gamma \, \hat{a}\right)\eta \left(\hat{a}^{\dagger }\right)\exp \left(\gamma \, \hat{a}\right)=\eta \left(\hat{a}^{\dagger
}-\gamma \right),
\end{equation}
 for any well behaved function \(\eta \left(\hat{a}^{\dagger }\right)\); thus
\begin{equation}
|\frac{x}{\sqrt{2}};r\rangle =\frac{1}{\sqrt{\mu }}\exp \left(\frac{x^2}{4}\right)\exp \left[\frac{x}{\sqrt{2}}\left(\hat{a}^{\dagger }-\frac{x}{\sqrt{2}}\right)\right]\exp
\left[-\frac{\nu }{2\mu }\left(\hat{a}^{\dagger }-\frac{x}{\sqrt{2}}\right)^2\right]|0\rangle .
\end{equation}
After some algebra,
\begin{equation}
|\frac{x}{\sqrt{2}};r\rangle =\frac{1}{\sqrt{\mu }}\exp \left[-\frac{x^2}{4}\left(1+\frac{\nu }{\mu }\right)\right]\exp \left[-\frac{\nu }{2\mu
}\hat{a}^{\dagger^{2}}+\frac{x}{\sqrt{2}}\left(1+\frac{\nu }{\mu }\right)\hat{a}^{\dagger }\right]|0\rangle .
\end{equation}
We take now the limit when \(r\to \infty\), or \(\frac{\nu }{\mu }\to 1\), so
\begin{equation}
|x\rangle _p\propto \exp \left(-\frac{x^2}{2}\right)\exp \left(-\frac{\hat{a}^{\dagger^{2} }}{2}+\sqrt{2}x\hat{a}^{\dagger }\right)|0\rangle .
\end{equation}
We get an expression that gives us the position eigenstates as an operator applied to the vacuum. Unlike the Yuen case, expression (21), now we have an \(x\) dependence and it looks like a better candidate to be the position eigenstate. In fact, in the next Section, we will show that this really
is an eigenstate of the position.

\section{Leaving squeezed states aside}
We will try now an alternative approach to the eigenstates of the position. We can write a position eigenstate, simply by multiplying it by a proper unit operator
\begin{equation}
|x\rangle_{p}=\sum_{n=0}^{\infty}\left| n\right\rangle \left\langle n|x\right\rangle_{p}
\end{equation}
Therefore the position eigenstate \(|x\rangle _p\) may be written as [13]
\begin{equation}
|x\rangle _p=\sum _{n=0}^{\infty } \psi _n(x)|n\rangle
\end{equation}
with \(\psi _n(x)=\frac{1}{\sqrt{2^n\sqrt{\pi }n!}}e^{\left.-x^2\right/2}H_n(x)\); such that \(|x\rangle _p\) may be re-written as
\begin{equation}
|x\rangle _p=\frac{e^{\left.-x^2\right/2}}{\pi ^{1/4}}\sum _{n=0}^{\infty } \frac{1}{2^{n/2}n!}H_n(x)\hat{a}^{\dagger^{n} }|0\rangle ,
\end{equation}
that may be added via using the generating function for Hermite polynomials [14]
\begin{equation}
e^{-t^2+2t\, x}=\sum _{n=0}^{\infty } H_k(x)\frac{t^k}{k!},
\end{equation}
to give
\begin{equation}
|x\rangle _p=\frac{e^{\left.-x^2\right/2}}{\pi ^{1/4}}e^{-\frac{\hat{a}^{\dagger^{2} }}{2}+\sqrt{2}x \hat{a}^{\dagger }}|0\rangle .
\end{equation}
The above expression allows us to write the position eigenstate as an operator applied to the vacuum. Note that this expression is the same as the one obtained using the Caves definition for the squeezed states, formula (28). We prove now that indeed (32) is an eigenvector of the postion operator; for that, we write the position operator as \(\hat{x}=\frac{\hat{a}+\hat{a}^{\dagger }}{\sqrt{2}}\), thus
\begin{equation}
\hat{x}|x\rangle _p=\frac{e^{\left.-x^2\right/2}}{\pi ^{1/4}\sqrt{2}}\left(\hat{a}+\hat{a}^{\dagger }\right)e^{-\frac{\hat{a}^{\dagger^{2} }}{2}+\sqrt{2}x\hat{a}^{\dagger }}|0\rangle .
\end{equation}
Inserting the identity operator in the above expression as \(\hat{I}=e^{-\frac{\hat{a}^{\dagger^{2} }}{2}}e^{\sqrt{2}x \hat{a}^{\dagger }}e^{-\sqrt{2}x \hat{a}^{\dagger }}e^{\frac{\hat{a}^{\dagger^{2} }}{2}}\), we get
\begin{equation}
\hat{x}|x\rangle _p=\frac{e^{\left.-x^2\right/2}}{\pi ^{1/4}\sqrt{2}}e^{-\frac{\hat{a}^{\dagger^{2} }}{2}}e^{\sqrt{2}x \hat{a}^{\dagger }}e^{-\sqrt{2}x
\hat{a}^{\dagger }}e^{\frac{\hat{a}^{\dagger^{2} }}{2}}\left(\hat{a}+\hat{a}^{\dagger }\right)e^{-\frac{\hat{a}^{\dagger^{2} }}{2}}e^{\sqrt{2}x \hat{a}^{\dagger}}|0\rangle ;
\end{equation}
as \(e^{\frac{\hat{a}^{\dagger^{2} }}{2}}\left(\hat{a}+\hat{a}^{\dagger }\right)e^{-\frac{\hat{a}^{\dagger^{2} }}{2}}=\hat{a}-\hat{a}^{\dagger }+\hat{a}^{\dagger
}=\hat{a}\), \(e^{\frac{\hat{a}^{\dagger^{2} }}{2}}\left(\hat{a}+\hat{a}^{\dagger }\right)e^{-\frac{\hat{a}^{\dagger^{2} }}{2}}=\hat{a}-\hat{a}^{\dagger
}+\hat{a}^{\dagger }=\hat{a}\), and \(\hat{a}|0\rangle =0\), we obtain
\begin{equation}
\hat{x}|x\rangle _p=x\frac{e^{\left.-x^2\right/2}}{\pi ^{1/4}}e^{-\frac{\hat{a}^{\dagger^{2} }}{2}}e^{\sqrt{2}x \hat{a}^{\dagger }}|0\rangle =x|x\rangle_p,
\end{equation}
as we wanted to show.

We can write (32) in terms of coherent states. We have
\begin{equation}
e^{\sqrt{2}x \hat{a}^{\dagger }}|0\rangle =\sum _{k=0}^{\infty } \frac{1}{k!}\left(\sqrt{2}x\right)^k\hat{a}^{\dagger^{k} }|0\rangle =\sum _{k=0}^{\infty
} \frac{\left(\sqrt{2}x\right)^k}{\sqrt{k!}}|k\rangle =e^{x^2}|\sqrt{2}x\rangle ,
\end{equation}
thus
\begin{equation}
|x\rangle _p=\frac{e^{\left.x^2\right/2}}{\pi ^{1/4}}e^{-\frac{\hat{a}^{\dagger^{2} }}{2}}|\sqrt{2}x\rangle .
\end{equation}

With the expressions obtained, it is easy to show that the squeezed states have the form of a Gaussian wave packet. To confirm this, we use the above expression to state that
\begin{equation}
\langle \alpha ;r|x\rangle _p=\langle \alpha |\hat{S}^{\dagger }(r)|x\rangle _p=\frac{e^{\left.x^2\right/2}}{\pi ^{1/4}}\langle \alpha |\hat{S}^{\dagger
}(r)e^{-\frac{\hat{a}^{\dagger^{2} }}{2}}|\sqrt{2}x\rangle .
\end{equation}
We write \(\hat{S}^{\dagger }(r)e^{-\frac{\hat{a}^{\dagger^{2} }}{2}}\) as \(e^{-\frac{\hat{a}^{\dagger^{2} }}{2}}e^{\frac{\hat{a}^{\dagger^{2} }}{2}}\hat{S}^{\dagger
}(r)e^{-\frac{\hat{a}^{\dagger^{2} }}{2}}\), where we have just inserted the identity operator \(\hat{I}=e^{-\frac{\hat{a}^{\dagger^{2} }}{2}}e^{\frac{\hat{a}^{\dagger^{2}
}}{2}}\), and we use that \(e^{\frac{\hat{a}^{\dagger }}{2}}\eta \left(\hat{a}\right)e^{-\frac{\hat{a}^{\dagger }}{2}}=\eta \left(\hat{a}-\hat{a}^{\dagger
}\right)\), for any well behaved function \(\eta\), to obtain
\begin{equation}
\langle \alpha ;r|x\rangle _p=\frac{\exp \left[\frac{1}{2}\left(x^2-r\right)\right]}{\pi ^{1/4}}\langle \alpha |e^{-\frac{\hat{a}^{\dagger^{2} }}{2}}e^{\frac{r}{2}\hat{a}^2-r\hat{a}^{\dagger
}\hat{a}}|\sqrt{2}x\rangle .
\end{equation}
As the coherent states \(|\alpha \rangle\) are eigenfunctions of the annihilation operator \(\hat{a}\), it is very easy to show that \(\langle \alpha|e^{-\frac{\hat{a}^{\dagger^{2} }}{2}}=\langle \alpha |e^{-\frac{\alpha ^{*^{2}}}{2}}\), so
\begin{equation}
\langle \alpha ;r|x\rangle _p=\frac{\exp \left[\frac{1}{2}\left(x^2-\alpha^{*^{2}} -r\right)\right]}{\pi ^{1/4}}\langle \alpha |e^{\frac{r}{2}\hat{a}^2-r\hat{a}^{\dagger}\hat{a}}|\sqrt{2}x\rangle .
\end{equation}
In the Appendix, we disentangle the operator \(e^{\frac{r}{2}\hat{a}^2-r\hat{a}^{\dagger }\hat{a}}\) as \(e^{-r\hat{a}^{\dagger }\hat{a}}e^{\frac{1-e^{2r}}{4}\hat{a}^2}\),
and we get
\begin{equation}
\langle \alpha ;r|x\rangle _p=\frac{\exp \left[\frac{1}{2}\left(x^2-\alpha^{*^{2}} -r\right)\right]}{\pi ^{1/4}}\langle \alpha |e^{-r\hat{a}^{\dagger
}\hat{a}}e^{\frac{1-e^{2r}}{4}\hat{a}^2}|\sqrt{2}x\rangle .
\end{equation}
It is very easy to see that \(e^{\frac{1-e^{2r}}{4}\hat{a}^2}|\sqrt{2}x\rangle =e^{\frac{1-e^{2r}}{2}x^2}|\sqrt{2}x\rangle\), and that \(e^{i \gamma
 \hat{n}}|\alpha \rangle =\left|e^{i \gamma }\alpha \right\rangle\), thus
\begin{equation}
\left\langle \alpha;r|x  \right\rangle _{p}=\frac{1}{\pi^{1/4}}\exp\left\lbrace\dfrac{1}{2}\left[ \left( 2-e^{2r}\right)x^{2}-{\alpha^{*}}^{2}-r  \right]  \right\rbrace  \left\langle \alpha|\sqrt{2}e^{-r}x\right\rangle .
\end{equation}
Finally, as \(\langle \delta |\epsilon \rangle =e^{-\frac{1}{2}\left(\left| \delta \right| ^2+\left| \epsilon \right| ^2-2\delta ^*\epsilon \right)}\),
we have
\begin{equation}
\langle \alpha ;r|x\rangle _p=\frac{1}{\pi ^{1/4}}\exp \left\{\frac{1}{2}\left[\left(2-e^{2r}-2e^{-2r}\right)x^2+2\sqrt{2}\alpha ^*e^{-r}x-\alpha^{*^{2}}
-\left| \alpha \right| ^2-r\right]\right\},
\end{equation}
as we wanted to show.

\section{The Husimi $\mathcal{Q}$-function}
We can now find the wave function of a coherent state as a function of the position [15]. We use equation (32), that express the eigenstates of the position as an operator acting on the vacuum, and get that
\begin{equation}
\left\langle \beta|x\right\rangle _{p}=\frac{e^{-x^{2}/2}}{\pi^{1/4}}\left\langle \beta|e^{-\frac{\hat{a}^{\dagger^{2} }}{2} +\sqrt{2} x \hat{a}^{\dagger }}|0\right\rangle = 
\frac{e^{-\frac{x^2}{2}}}{\sqrt[4]{\pi }}
e^{-\frac{\beta^{*^{2}} }{2}+\sqrt{2} \beta ^* x} \left\langle \beta|0\right\rangle 
=\frac{e^{-\frac{x^2}{2}}}{ \pi^{1/4}  }
e^{-\frac{\beta^{*^{2}} }{2}-\frac{\left| \beta \right| ^2}{2}+\sqrt{2} \beta ^* x}
\end{equation}
as \(\langle \beta |\hat{a}^{\dagger }=\beta ^*\langle \beta |\) and \(\langle \beta |n\rangle =e^{-\frac{\left|\beta |^2\right.}{2}}\frac{\beta^{*^{n}}
}{\sqrt{n!}}\). 

The Husimi $\mathcal{Q}$-function [16] can be calculated from (45) simply as
\begin{equation}
\mathcal{Q}(\beta)=\frac{1}{\pi}
\left|\left\langle\beta|x\right\rangle _{p}  \right| ^{2}=
 \frac{e^{-x^2} e^{-\left| \beta \right| ^2}}{\pi ^{3/2}}
\left| e^{-\frac{\beta^{*^{2}}}{2}+\sqrt{2} \beta ^* x}    \right| ^2
\end{equation}
that after some algebra, can be re-written as
\begin{equation}
\mathcal{Q}(\beta)=\frac{1}{\pi ^{3/2}}  \exp \left[ -x^2 - \left| \beta \right|^2    
-\text{Re}({\beta^*}^2)+2\sqrt{2}\text{Re}(\beta)x\right] 
\end{equation}

In Figures 1 and 2, we plot the Husimi $\mathcal{Q}$-function for different values of \(x\).
\begin{figure}[h]
\centering
\includegraphics[width=0.4\linewidth]{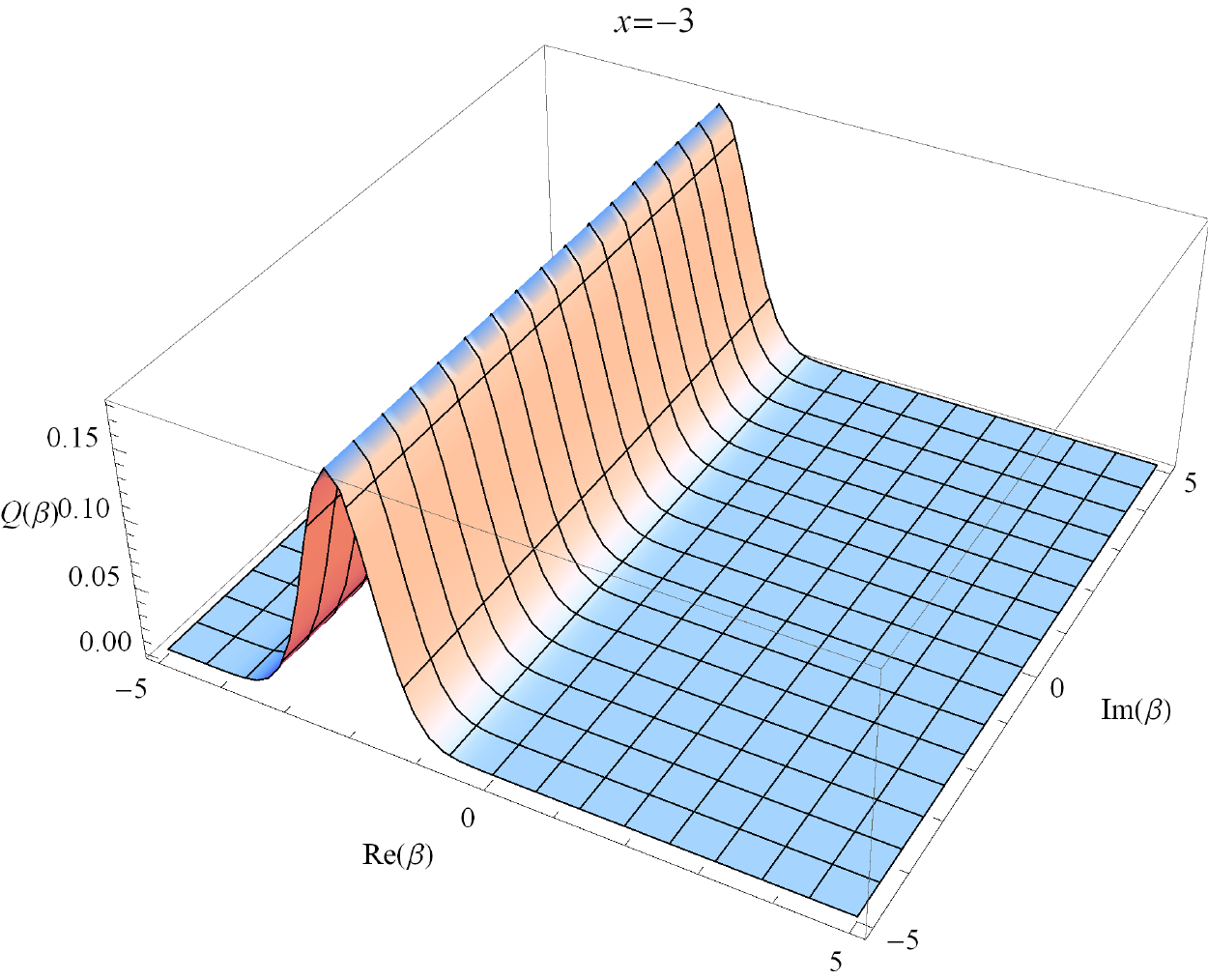}
\qquad \qquad
\includegraphics[width=0.4\linewidth]{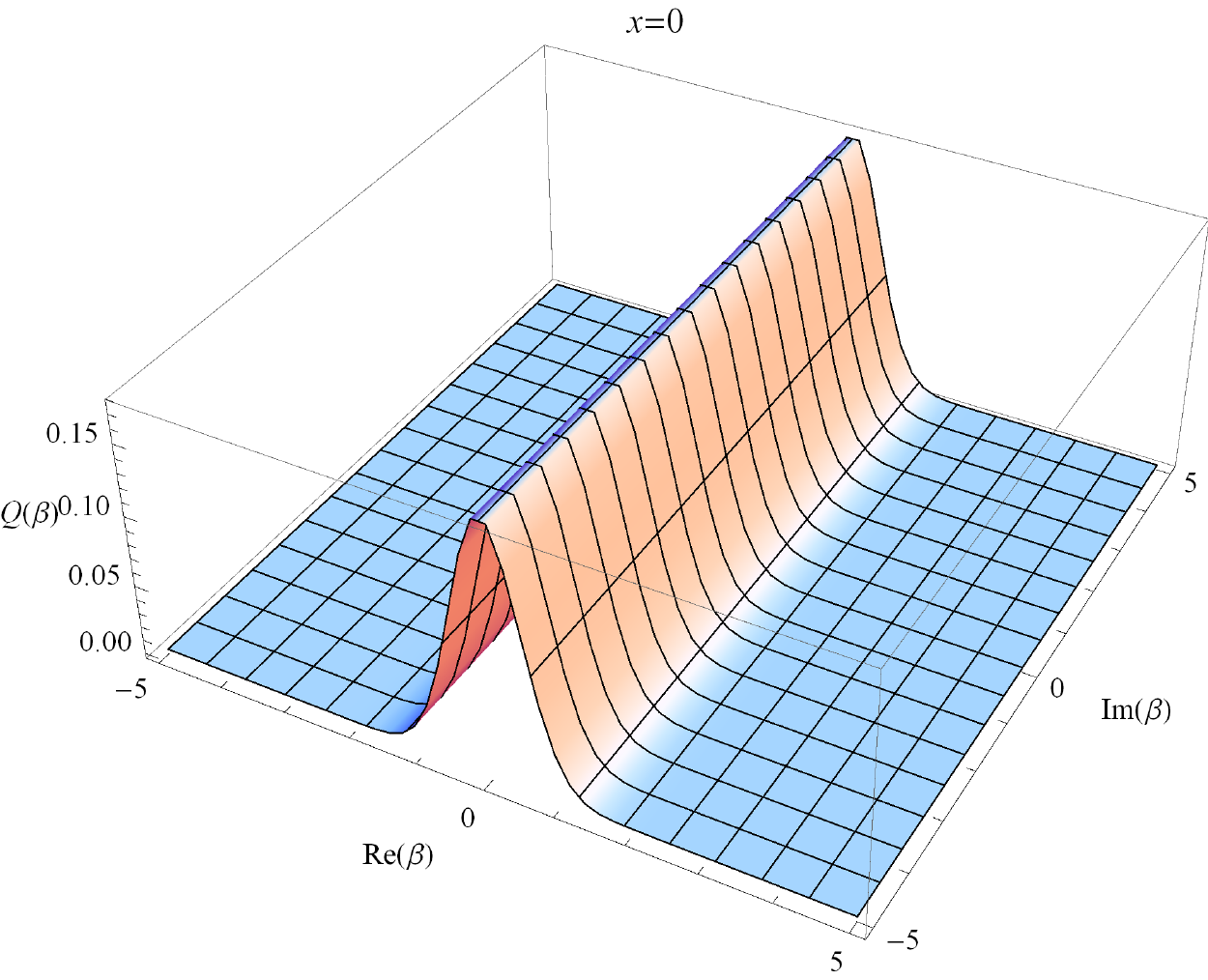}
\caption{The Husimi $\mathcal{Q}$-function for $x=-3$ and for $x=0$.}
\end{figure}
\begin{figure}[h]
\centering
\includegraphics[width=0.4\linewidth]{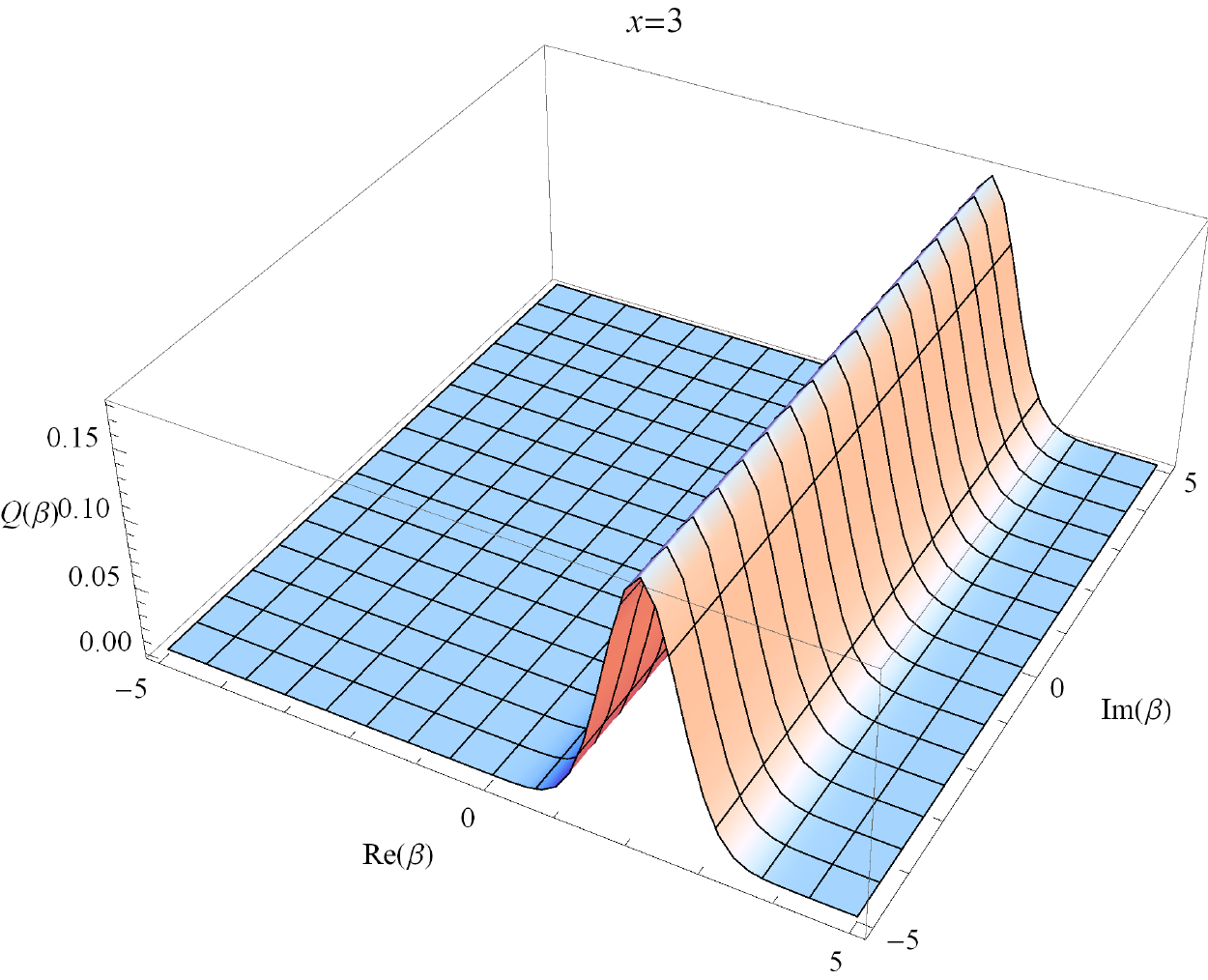}
\qquad \qquad
\includegraphics[width=0.4\linewidth]{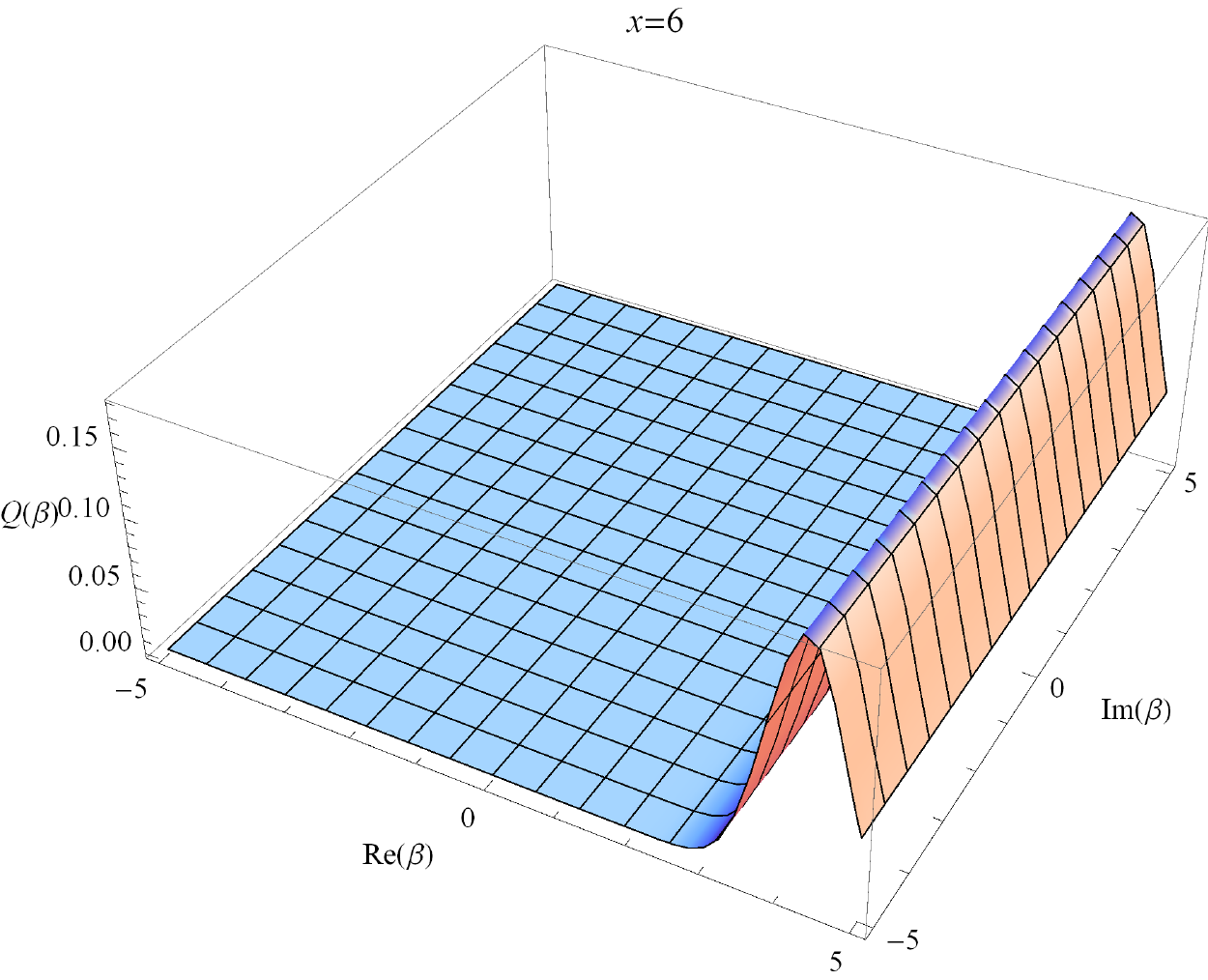}
\caption{The Husimi $\mathcal{Q}$-function for $x=3$ and for $x=6$.}
\end{figure}

\section{Conclusions}
We have found an operator that applied to the vacuum gives us the eigenstates of the position. We did that by two ways; first, using the Caves definition of the squeezed states, we took the limit of extreme squeezing in the position side, to get the position eigenstate. Second, we used the expansion of an arbitrary wave function in the base of the harmonic oscillator; i.e., we wrote an arbitrary wave function in terms of Hermite polynomials.
The expressions obtained allows us to show certain properties of the squeezed states, and also allow us to write in a very easy way the Husimi \textit{Q}-function of the position eigenstates. The same procedure can be followed to find the eigenstates of the momentum, but taken the limit when the squeeze parameters goes to \(-\infty\).\\
We can also conclude that from the point of view of this work, the Caves approach to the squeeze states is more adequate, since it gives the correct eigenstates of the position; while the Yuen definition, formula (1), gives an expression that is incorrect. So, we must first squeeze the vacuum, and after, we displace it.

\appendix
\section{Appendix}
In this appendix, we show how to disentangle the operator \(e^{-\frac{r}{2}\hat{a}^2+r\hat{a}^{\dagger }\hat{a}}\). We define
\begin{equation}
\hat{F}(r)\equiv e^{-\frac{r}{2}\hat{a}^2+r\hat{a}^{\dagger }\hat{a}},
\end{equation}
and we suppose that (48) can be rewritten as
\begin{equation}
\hat{F}(r)=\exp \left[f(r)\hat{a}^{\dagger }\hat{a}\right]\exp \left[g(r)\hat{a}^2\right],
\end{equation}
where \(f(r)\) and \(g(r)\) are two unknown well behaved functions; as \(\hat{F}(0)=\hat{I}\), being \(\hat{I}\) the identity operator, these functions
most satisfy the conditions \(f(0)=g(0)=0\). At first sight, one can think that in the proposal (45) should be a term of the form \(\exp \left[h(r)\hat{a}^{\dagger^{2}
}\right]\); however, this is not the case because \(\left[\hat{a}^2,\hat{a}^{\dagger }\hat{a}\right]=2\hat{a}^2\).
We differentiate with respect to \(r\), to find
\begin{equation}
\frac{d\hat{F}}{dr}=\frac{df}{dr}\hat{a}^{\dagger }\hat{a}\exp \left[f\hat{a}^{\dagger }\hat{a}\right]\exp \left[g\hat{a}^2\right]+\frac{dg}{dr}\exp
\left[f\hat{a}^{\dagger }\hat{a}\right]\hat{a}^2\exp \left[g\hat{a}^2\right],
\end{equation}
where for simplicity in the notation, we have dropped all \(r\)-dependency; we write the identity operator as \(\hat{I}=\exp \left[-f\hat{a}^{\dagger
}\hat{a}\right]\exp \left[f\hat{a}^{\dagger }\hat{a}\right]\) in the second term, to obtain
\begin{equation}
\frac{d\hat{F}}{dr}=\frac{df}{dr}\hat{a}^{\dagger }\hat{a}\exp \left[f\hat{a}^{\dagger }\hat{a}\right]\exp \left[g\hat{a}^2\right]+\frac{dg}{dr}\exp
\left[f\hat{a}^{\dagger }\hat{a}\right]\hat{a}^2\exp \left[-f\hat{a}^{\dagger }\hat{a}\right]\exp \left[f\hat{a}^{\dagger }\hat{a}\right]\exp \left[g\hat{a}^2\right].
\end{equation}
Using the Hadamard\'{ }s lemma [10, 11], it is very easy to prove that
\begin{equation}
\exp \left[f\hat{a}^{\dagger }\hat{a}\right]\hat{a}^2\exp \left[-f\hat{a}^{\dagger }\hat{a}\right]=e^{-2f}\hat{a}^2,
\end{equation}
so
\begin{equation}
\frac{d\hat{F}}{dr}=\left(\frac{df}{dr}\hat{a}^{\dagger }\hat{a}+\frac{dg}{dr}e^{-2f}\hat{a}^2\right)\hat{F}.
\end{equation}
Equating this equation to the one obtained differentiating the original formula for \(\hat{F}(r)\), equation (44), we get the following system of
first order ordinary differential equations
\begin{equation}
\frac{df}{dr}=1,\text{  }\frac{dg}{dr}e^{-2f}=-\frac{1}{2}
\end{equation}
The solution of the first equation, that satisfies the initial condition \(f(0)=0\), is the function \(f(r)=r\). Substituting this solution in the
second equation and solving it with the initial condition \(g(0)=0\), we obtain \(g(r)=\frac{1-e^{2r}}{4}\). Thus, finally we write
\begin{equation}
e^{-\frac{r}{2}\hat{a}^2+r\hat{a}^{\dagger }\hat{a}}=e^{r\hat{a}^{\dagger }\hat{a}}e^{\frac{1-e^{2r}}{4}\hat{a}^2}
\end{equation}

\end{document}